\documentclass[conference]{IEEEtran}

\usepackage{amsmath, amssymb}
\usepackage{graphicx}
\usepackage{subcaption}
\usepackage{booktabs} 
\usepackage{tabularx}  
\usepackage{cite} 
\usepackage{algorithm}
\usepackage{algpseudocode}
\usepackage{url}
\usepackage{balance}
\usepackage{xcolor}
\usepackage[switch]{lineno} 
\usepackage{authblk}

\allowdisplaybreaks

\usepackage[bookmarks=false]{hyperref}

\hypersetup{
    colorlinks=true,
    linkcolor=blue,
    filecolor=magenta,      
    urlcolor=cyan,
    citecolor=black,
    pdftitle={}, 
    pdfpagemode=FullScreen,
    }
\DeclareMathOperator*{\argmax}{arg\,max}

\begin{document}
\setlength{\textheight}{9.25in}
\setlength{\columnsep}{0.21in}

\nolinenumbers 

\title{Multi-Agent Deep Reinforcement Learning for Collaborative UAV Relay Networks under Jamming Atatcks}

\author[1]{Thai Duong Nguyen}
\author[1]{Ngoc-Tan Nguyen}
\author[1]{Thanh-Dao Nguyen}
\author[2]{Nguyen Van Huynh}
\author[3]{\authorcr Dinh-Hieu Tran}
\author[3]{Symeon Chatzinotas}

\affil[1]{\textit{VNU - University of Engineering and Technology (VNU - UET), Hanoi, Vietnam}}
\affil[2]{\textit{School of Computer Science and Informatics, University of Liverpool, United Kingdom}}
\affil[3]{\textit{Interdisciplinary Centre for Security Reliability and Trust (SnT), University of Luxembourg, Luxembourg}}

\maketitle

\begin{abstract}
The deployment of Unmanned Aerial Vehicle (UAV) swarms as dynamic communication relays is critical for next-generation tactical networks. However, operating in contested environments requires solving a complex trade-off, including maximizing system throughput while ensuring collision avoidance and resilience against adversarial jamming. Existing heuristic-based approaches often struggle to find effective solutions due to the dynamic and multi-objective nature of this problem. This paper formulates this challenge as a cooperative Multi-Agent Reinforcement Learning (MARL) problem, solved using the Centralized Training with Decentralized Execution (CTDE) framework. Our approach employs a centralized critic that uses global state information to guide decentralized actors which operate using only local observations. Simulation results show that our proposed framework significantly outperforms heuristic baselines, increasing the total system throughput by approximately 50\% while simultaneously achieving a near-zero collision rate. A key finding is that the agents develop an emergent anti-jamming strategy without explicit programming. They learn to intelligently position themselves to balance the trade-off between mitigating interference from jammers and maintaining effective communication links with ground users.
\end{abstract}

\begin{IEEEkeywords} 
Multi-Agent Deep Reinforcement Learning (MADRL), Centralized Training with Decentralized Execution (CTDE), UAV, Trajectory Planning, Resource Allocation.
\end{IEEEkeywords}

\section{Introduction}
\label{sec:introduction}

Unmanned Aerial Vehicles (UAVs) have emerged as a cornerstone of modern network architectures, poised to provide agile and on-demand connectivity for next-generation systems, ranging from 6G cellular networks to tactical edge computing \cite{Zeng2019_ProcIEEE}. In military and disaster-response scenarios, their ability to rapidly deploy as aerial base stations for ground assets, such as Ground Combat Vehicles (GCVs), is transformative. However, these deployments face considerable challenges, including both physical and spectral issues. In particular, a swarm of UAVs must not only navigate complex terrains and avoid collisions but also withstand complex electronic warfare, such as adversarial jamming, which can cripple communication links and compromise mission objectives \cite{Pirayesh2023_COMST_Jamming}. This creates a fundamental trade-off between maximizing communication performance and ensuring multi-faceted operational survivability.

Prior research has typically addressed these challenges from two main perspectives. One significant body of work has focused on communication-centric optimization, employing mathematical programming to determine optimal static placements for UAVs that maximize throughput or coverage \cite{Mozaffari2019_ComST_Survey}. While theoretically sound, this static paradigm creates predictable and high-value targets, rendering the network highly vulnerable to kinetic or electronic attacks. Another line of research has pursued survivability through motion, proposing pre-computed or reactive trajectories to evade threats \cite{EnergyEff_Zeng2017}. Yet, such approaches are often rigid, failing to adapt in real-time to the unpredictable mobility of allied ground units or the dynamic nature of emergent threats. The fundamental limitation of these prior works is the tendency to treat these challenges as separable sub-problems. This overlooks the deeply coupled nature of an agent's physical positioning, its resource allocation, and the overall spectral resilience of the network.

This paper argues that the key to resilient UAV networks lies in learning an optimal \textit{behavioral policy} rather than finding an optimal \textit{position}. We propose that robust operational patterns should be an \textit{emergent property} of a holistic and multi-objective learning process. Accordingly, we formulate the problem within the framework of Multi-Agent Reinforcement Learning (MARL)\cite{SuttonBarto_Book}. We model the UAV swarms as a cooperative team of intelligent agents tasked with learning a complex and decentralized policy that dynamically co-optimizes network performance and survivability in response to real-time environmental feedback. Our approach utilizes the Centralized Training with Decentralized Execution (CTDE) architecture, which is well-suited for this task \cite{lowe2020multiagentactorcriticmixedcooperativecompetitive}. It enables the team to learn complex cooperative strategies from global information during an offline training phase, while executing actions based solely on local observations during deployment, thereby removing the need for high-bandwidth inter-agent communication in the field.

The primary contributions of this work are threefold:
\begin{itemize}
    \item We present a comprehensive MARL formulation for the joint UAV operation problem that integrates throughput maximization, collision avoidance, and jamming resilience into a unified reward structure.
    \item We propose and implement a CTDE-based learning framework that employs a centralized critic and decentralized actors to learn complex cooperative policies for resource allocation and spatio-temporal positioning.
    \item We provide extensive simulation results demonstrating that the proposed agents learn emergent anti-jamming strategies and effectively balance the trade-off between throughput and safety, significantly outperforming established heuristic baselines.
\end{itemize}

The remainder of this paper is organized as follows. Section \ref{sec:SysModelAndProblem} details the system model and formalizes the problem. Section \ref{sec:solution} presents our proposed MARL-based solution. Section \ref{sec:PerfEval} provides an in-depth analysis and discussion of the simulation results, and section \ref{sec:Conclusion} concludes the paper with insights into future research directions.

\section{System Model and Problem Formulation}
\label{sec:SysModelAndProblem}

\begin{figure}[htbp] 
    \centering
    \includegraphics[width=1\columnwidth]{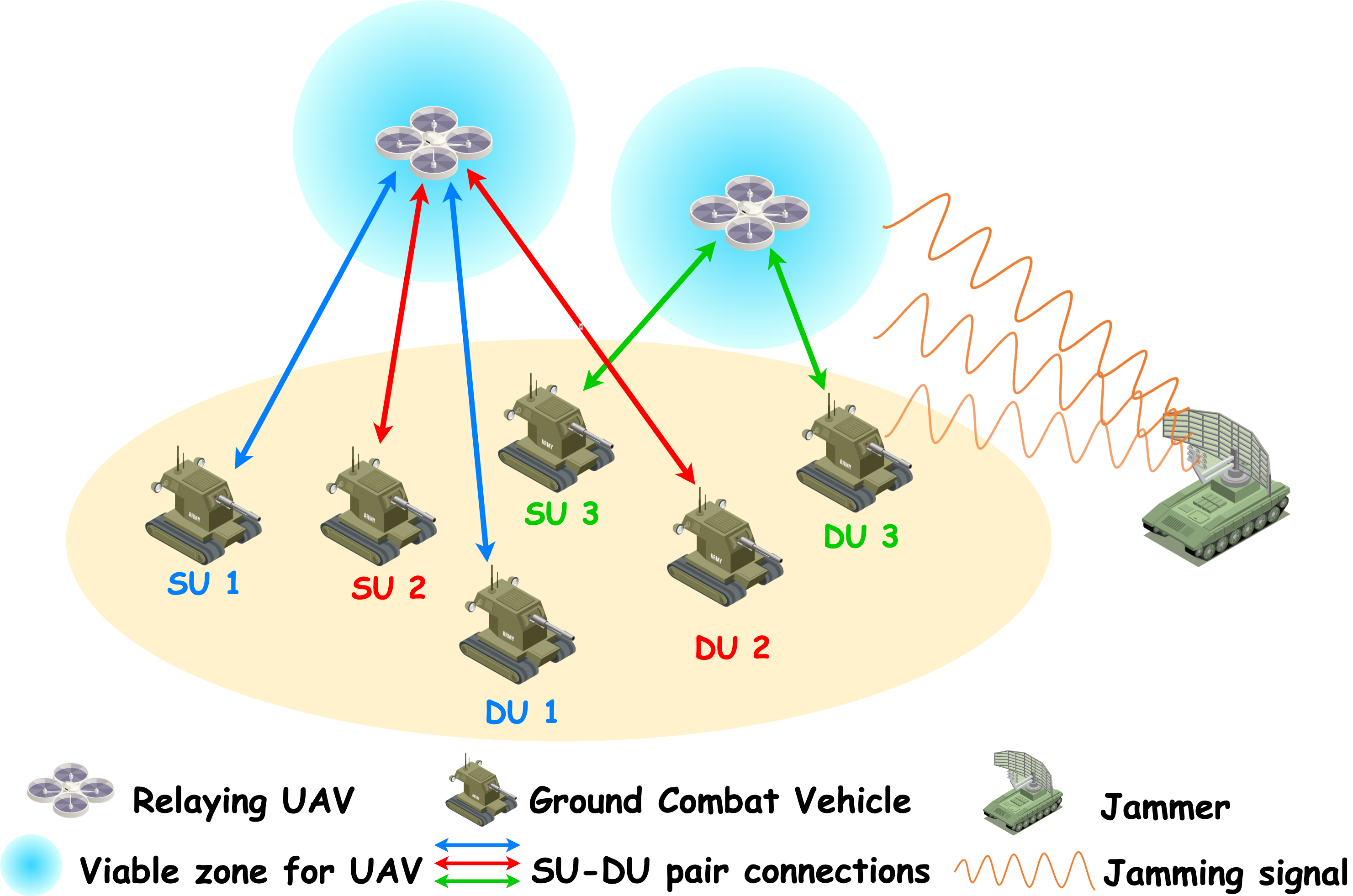} 
    \caption{System model}
    \label{fig:system_model}
    \vspace{-3mm} 
\end{figure}

To ground our investigation in a realistic operational context, this section first establishes the network architecture, the physical environment, and the communication channel characteristics. We then formalize the multi-faceted challenge of resilient network operation as a cooperative multi-agent reinforcement learning problem in Fig.~\ref{fig:system_model}

\subsection{Network and Environment Model}
We define a three-dimensional operational theater consisting of a $120 \times 120$~m ground area, where UAVs operate at altitudes between 15~m and 35~m. The system comprises three key entity types:

\begin{itemize}
    \item \textbf{UAV Agents ($\mathcal{U}$):} A set of $N_U$ autonomous Unmanned Aerial Vehicles, $\mathcal{U} = \{U_1, \dots, U_{N_U}\}$, acts as the core of our dynamic communication infrastructure. Each agent $U_i$ is tasked with serving ground users while navigating the environment and managing its own state, including its 3D position $p_i(t) \in \mathbb{R}^3$ and its remaining battery energy $E_i(t)$.

    \item \textbf{Ground Users ($\mathcal{C}$):} A set of $N_C$ ground combat vehicle pairs (GCVs), $\mathcal{C} = \{C_1, \dots, C_{N_C}\}$, where each pair $C_n$ consists of a source user $S_n$ and a destination user $D_n$. These users move with unpredictable mobility patterns on the ground and require a two-hop communication link relayed by the UAVs. In our experimental setup, we consider a specific configuration with an equal number of agents and user pairs, setting $N_U = N_C = 5$.

    \item \textbf{Adversarial Jammers ($\mathcal{J}$):} The operational environment is contested by a set of $J$ static jammers, $\mathcal{J} = \{J_1, \dots, J_J\}$, located at fixed positions $p_j^{\text{jam}} \in \mathbb{R}^3$. These jammers continuously emit interference signals with a constant power $P^{\text{jam}}$, representing the primary spectral threat to the network's stability.
\end{itemize}
The system evolves in discrete time steps, where at each step, UAV agents observe the environment, select actions, and receive feedback.

\subsection{Communication and Interference Model}

We model the channel gain between nodes at positions $p_x$ and $p_y$ using a standard path loss model, $G_{x,y} = \|p_x - p_y\|^{-\alpha}$, with the path loss exponent $\alpha=2.0$ \cite{Goldsmith2005_Book}. While this deterministic model does not capture more complex, probabilistic effects common in UAV-to-ground links such as Rician fading, it is a foundational approach widely adopted in network-level optimization studies. This simplification allows our analysis to focus squarely on the emergent, strategic behaviors of the MARL agents, rather than the stochasticity of the physical layer \cite{Zeng2019_ProcIEEE, EnergyEff_Zeng2017}.

For a communication link from a transmitter $x$ to a receiver $y$, the Signal-to-Interference-plus-Noise Ratio (SINR) is given by:
\begin{equation}
\label{eq:sinr}
\gamma_{x,y} = \frac{P_x \cdot G_{x,y}}{\sigma^2 + I_{\text{co}}(y) + I_{\text{jam}}(y)},
\end{equation}
where $P_x$ is the transmit power of node $x$ and $\sigma^2$ is the thermal noise power. The interference at receiver $y$ comprises two components:
\begin{itemize}
    \item $I_{\text{co}}(y)$: Co-channel interference generated by other friendly transmitters in the network operating concurrently.
    \item $I_{\text{jam}}(y) = \sum_{j=1}^{J} P^{\text{jam}} \cdot G_{j,y}$: The cumulative jamming interference from all adversarial jammers.
\end{itemize}
The end-to-end data rate for a user pair $C_n$ relayed by a UAV is determined by the bottleneck of its two-hop link, calculated as $R_n = \min(R_{n,\text{ul}}, R_{n,\text{dl}})$, where the uplink from the source user to the UAV relay ($R_{n,\text{ul}}$) and the downlink from the UAV to the destination user ($R_{n,\text{dl}}$) rates are derived from their respective SINR values via the Shannon-Hartley theorem \cite{Shannon1948}.

\subsection{Problem Formulation as a DEC-POMDP}
We formulate the challenge of learning adaptive operational policies as a Decentralized Partially Observable Markov Decision Process (DEC-POMDP), which provides a natural mathematical framework for cooperative multi-agent decision-making under uncertainty \cite{Bernstein2002_Complexity}. The DEC-POMDP is defined by the tuple $(\mathcal{S}, \{\mathcal{A}_i\}, P, \{\mathcal{R}_i\}, \{\Omega_i\}, O, \gamma)$ with $O(o|s', a)$ is the observation function, defining the probability of receiving a joint observation $o$ given the resulting state $s'$.

\begin{itemize}
    \item \textbf{State Space ($\mathcal{S}$):} The global state $s_t \in \mathcal{S}$ encompasses the complete system information, including the positions, remaining energy, and stationary time ($\tau_{\text{stay}}$) of all UAVs, as well as the positions of all ground users. This global view is accessible only during the centralized training phase.

    \item \textbf{Observation Space ($\Omega$):} The local observation $o_{i,t} \in \Omega_i$ for each agent is composed of its internal state (position, energy, $\tau_{\text{stay}}$) and its perception of the local environment, specifically the relative positions and channel gains to its $K$-nearest ground users.

    \item \textbf{Action Space ($\mathcal{A}$):} The action space for each agent is discrete and identical, defined as the Cartesian product of a set of movement primitives and a set of transmission power levels, $\mathcal{A}_i = \mathcal{A}_{\text{move}} \times \mathcal{A}_{\text{power}}$. $\mathcal{A}_{\text{move}}$ contains 7 distinct movement vectors (Stay, North, South, East, West, Up, Down), and $\mathcal{A}_{\text{power}}$ contains 3 discrete power settings, resulting in 21 composite actions per agent.

    \item \textbf{Reward Function Design:} To guide the agents towards complex cooperative behaviors, we design a multi-objective reward structure. It is composed of two key components: a shared global reward ($R_t$) for learning uav-swarm's strategy, and a shaped individual reward ($r_{i,t}$) for providing agent-specific tactical guidance.
    
    \textbf{The Global Reward Function ($R_t$)} is a shared signal for all agents, designed to reflect the overall mission success. It is a weighted combination of system-wide objectives:
    \begin{equation}
    \label{eq:reward_global}
    R_t = w_{\text{thr}} \sum_{n=1}^{N_C} R_n(t) + w_{\text{coop}} R_{\text{coop}}(t) - \sum_{k \in \{\text{col}, \text{fly}\}} w_k P_k(t).
    \end{equation}
    where $\sum R_n(t)$ represents the total network throughput, $R_{\text{coop}}(t)$ denotes a cooperation bonus that rewards workload balance and desirable inter-agent spacing, and $P_{\text{col}}(t)$ and $P_{\text{fly}}(t)$ are penalties for collision risk and flight energy consumption, respectively \cite{DinhHieu2020CoarseTrajectory}. The weights $w_{(\cdot)}$ are dynamically adjusted during training via a homeostatic control mechanism, implementing a curriculum that progressively shifts the learning focus from safety to performance.

    \textbf{The Individual Reward Function ($r_{i,t}$)} is a dense, agent-specific signal used in the actor's learning update (as detailed in Section III-B). It is designed to provide more direct credit assignment by combining an agent's direct contribution to throughput with penalties and bonuses that shape its local behavior:
    \begin{equation}
    \label{eq:reward_individual}
     r_{i,t} \triangleq \alpha_{\text{thr}} \cdot R_{i,t} + \alpha_{\text{assign}} \cdot B_{\text{assign}}(i,t) - \alpha_{\text{col}} \cdot P_{\text{col}}(t),
    \end{equation}
    where $R_{i,t}$ is the throughput generated by agent $i$, $B_{\text{assign}}(i,t)$ is a bonus based on the number of GCVs assigned to agent $i$ to incentivize participation, and $P_{\text{col}}(t)$ is the shared collision penalty. The coefficients $\alpha_{(\cdot)}$ are hyperparameters that balance these individual objectives. This dual-reward structure is crucial for our learning algorithm, as it allows the critic to learn from the stable global signal while the actors receive more immediate, actionable feedback.
\end{itemize}

The objective is to find a set of decentralized policies $\{\pi_i(a_i|o_i)\}_{i=1}^N$ that maximizes the expected discounted cumulative global reward, $J(\pi) = \mathbb{E}_{\pi} \left[ \sum_{t=0}^{T} \gamma^t R_t \right]$, where $\gamma \in [0, 1)$ is the discount factor.

\section{Proposed MARL Framework}
\label{sec:solution}

To solve the formulated DEC-POMDP, we propose a MARL framework based on the CTDE paradigm \cite{amato2024introductioncentralizedtrainingdecentralized, lowe2020multiagentactorcriticmixedcooperativecompetitive}.

\subsection{CTDE Network Architecture}
\begin{figure}[htbp]
    \centering
    \includegraphics[width=0.95\columnwidth]{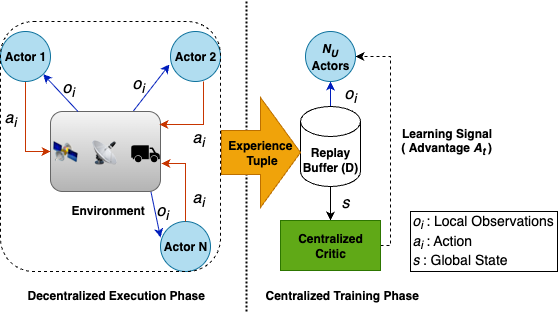}
    \caption{The architecture of the proposed CTDE framework.}
    \label{fig:ctde_architecture}
    \vspace{-3mm} 
\end{figure}

The overall architecture of our approach, illustrated in Fig.~\ref{fig:ctde_architecture}, separates the process into two distinct phases. During the decentralized execution phase, each agent's actor policy uses only local observations to select an action. Subsequently, in the centralized training phase, a shared critic leverages global state information from a replay buffer to effectively guide the policy updates for all actors, enabling the emergence of complex cooperative behaviors. The components of this architecture are detailed below.

To effectively address the multi-agent credit assignment problem and mitigate non-stationarity, we design a CTDE architecture comprising a centralized critic and multiple decentralized actors, with dedicated target networks to stabilize the learning process.

\begin{itemize}
    \item \textbf{Centralized Critic ($V_\phi$):} A single, centralized critic network, parameterized by $\phi$, is designed to learn the global state-value function, $V_\phi(s_t)$. Its sole input is the complete global state $s_t \in \mathcal{S}$, and its output is a scalar value estimating the expected discounted cumulative reward for the entire team from that state. By conditioning on the global state, the critic provides a stable and globally-aware learning signal to guide the training of all decentralized \textbf{actors}.


    \item \textbf{Decentralized Actors (Q-Networks):} Each agent's decentralized actor is a Q-network, parameterized by $\theta_i$, which maps the agent's local observation $o_{i,t}$ to a vector of action-values (Q-values), $Q_{\theta_i}(o_{i,t}, \cdot)$. The agent's policy $\pi_i$ is then derived from these values, typically by selecting the greedy action $a_t = \argmax_{a \in \mathcal{A}_i} Q_{\theta_i}(o_{i,t}, a)$ or by using an $\epsilon$-greedy exploration strategy.

    \item \textbf{Target Networks ($V_{\phi'}$ and $Q_{\theta'_i}$):} Following standard deep Q-learning techniques \cite{Mnih2015_Nature}, we maintain time-delayed copies of the critic and actor networks, denoted as $V_{\phi'}$ and $Q_{\theta'_i}$ respectively. These target networks are used to generate stable Temporal-Difference (TD) targets, a crucial practice for preventing training instability and divergence.
\end{itemize}

\subsection{Centralized Training with Advantage-Informed Q-Learning}
The core of our framework is an end-to-end training algorithm that synergizes principles from Double DQN \cite{vanHasselt2016_DoubleDQN} and advantage-based reward shaping, orchestrated through a prioritized experience replay mechanism to enhance learning efficiency \cite{Schaul2015_PER}. The process, detailed below, iterates over experiences sampled from a shared replay buffer $\mathcal{D}$.

\subsubsection{Experience Collection and Prioritized Replay}
During each episode, agents interact with the environment. The resulting transition tuples, containing the global state, local observations, actions, rewards, and next states $(s_t, \{o_{i,t}\}, \{a_{i,t}\}, R_t, \{r_{i,t}\}, s_{t+1}, \{o_{i,t+1}\})$, are stored in $\mathcal{D}$. We employ a prioritized replay buffer that samples experiences with a higher probability if they have a larger TD-error, allowing the model to focus on more informative transitions \cite{Schaul2015_PER}.

\subsubsection{Centralized Critic Update}
For a minibatch of transitions sampled from $\mathcal{D}$, the critic's parameters $\phi$ are updated to minimize the temporal-difference error. The TD target for the critic, $y_v$, is computed using the target critic network $V_{\phi'}$:
\begin{equation}
    y_v \triangleq R_t + \gamma V_{\phi'}(s_{t+1}).
\end{equation}
The critic is then updated by minimizing the importance-weighted Huber loss, which is robust to outliers \cite{Huber1964}:
\begin{equation}
    \mathcal{L}(\phi) = \mathbb{E}_{(s_t, R_t, s_{t+1}) \sim \mathcal{D}} \left[ w_j \cdot \text{Huber}(y_v - V_\phi(s_t)) \right],
\end{equation}
where $w_j$ is the importance sampling weight for sample $j$ to correct for the bias introduced by prioritized sampling.

\subsubsection{Decentralized Actor Update}
The update for each actor $\theta_i$ is where global information is injected into the local decision-making process. First, we compute the global advantage signal $A_t$ using both online and target critics:
\begin{equation}
\label{eq:advantage}
A_t \triangleq R_t + \gamma V_{\phi'}(s_{t+1}) - V_\phi(s_t).
\end{equation}
This advantage signal quantifies how much better or worse the team's collective action was compared to the expected outcome from state $s_t$.

Next, we construct the learning target $y_q$ for the actor's Q-function. To mitigate the overestimation bias common in Q-learning, we adopt the Double DQN approach \cite{vanHasselt2016_DoubleDQN}. The best next action is selected using the online actor network, but its value is evaluated using the target actor network:
\begin{equation}
    a'_{i, \text{best}} = \arg\max_{a' \in \mathcal{A}_i} Q_{\theta_i}(o_{i, t+1}, a'),
\end{equation}
\begin{equation}
    Q'_{\text{next}} = Q_{\theta'_i}(o_{i, t+1}, a'_{i, \text{best}}).
\end{equation}
The final TD target for the Q-value of the action $a_{i,t}$ taken at time $t$ is then a composite of the shaped \textbf{individual reward}, the scaled global advantage, and the discounted next-state Q-value:
\begin{equation}
\label{eq:q_target_final}
y_q \triangleq r_{i,t} + \lambda \cdot A_t + \gamma Q'_{\text{next}},
\end{equation}
where $r_{i,t}$ is the individual reward defined in Eq. (\ref{eq:reward_individual}) and $\lambda$ is a hyperparameter that controls the influence of the global advantage signal. This formulation, a key aspect of our approach, allows each agent to learn from its own specific reward signal while being steered by the team's overall performance. The actor's parameters $\theta_i$ are then updated by minimizing the following loss function:
\begin{equation}
\label{eq:actor_loss}
\mathcal{L}(\theta_i) = \mathbb{E}_{(s_t, o_{i,t}, a_{i,t}, \dots) \sim \mathcal{D}} \left[ w_j \cdot \text{Huber}(y_q - Q_{\theta_i}(o_{i,t}, a_{i,t})) \right].
\end{equation}
In this equation, the loss is the expected value of the Huber function applied to the TD-error, which is the discrepancy between the target Q-value ($y_q$) and the value predicted by the actor network ($Q_{\theta_i}$). The term $w_j$ represents the importance sampling weight used to correct for the bias introduced by the prioritized replay mechanism. We utilize the Huber loss for its robustness; it behaves quadratically for small errors similar to a mean-squared error loss, but linearly for large errors, making it less sensitive to noisy or outlier TD-error estimates and thus promoting more stable training.

\subsubsection{Target Network Updates}
Finally, the target network parameters $\phi'$ and $\theta'_i$ are updated periodically via a soft "Polyak" averaging rule, which promotes stable learning \cite{Polyak1992}:
\begin{equation}
\phi' \leftarrow \tau \phi + (1-\tau)\phi', \quad \theta'_i \leftarrow \tau \theta_i + (1-\tau)\theta'_i,
\end{equation}
where $\tau \ll 1$ is the update rate. After convergence, the trained, decentralized actor policies $\pi_i$ are deployed onto the UAVs for mission execution.

\section{Performance Evaluation}
\label{sec:PerfEval}

To validate the efficacy and elucidate the learned behaviors of our proposed MARL framework, we conduct a series of comprehensive simulations. This section details the experimental setup, defines the baseline policies used for comparison, and provides an in-depth analysis of the results.

\subsection{Experimental Setup}

We evaluate our framework in a simulated tactical environment where a team of UAV agents serves multiple GCV pairs under adversarial jamming. Each agent's policy is represented by a Multi-Layer Perceptron network. The training process is stabilized using online state normalization and guided by a curriculum learning strategy that progressively shifts the focus from safety- to performance-oriented behaviors. For reproducibility, all results are based on a single training run with a fixed random seed. The plotted curves are smoothed via a moving average for clarity, and all specific simulation and training hyperparameters are detailed in Table~\ref{tab:hyperparameters}.
\begin{table}[htbp] 
\centering
\caption{Key Simulation and Training Hyperparameters}
\label{tab:hyperparameters}
\begin{tabularx}{\columnwidth}{lX} 
\toprule 
\textbf{Parameter} & \textbf{Value} \\
\midrule 
\multicolumn{2}{c}{\textbf{Environment Configuration}} \\
\midrule
Operational Area Side ($L$) & 120 m \\
Altitude Range ($H_{\min}, H_{\max}$) & [15, 35] m \\
Number of UAVs ($N_U$) & 5 \\
Number of GCV Pairs ($N_C$) & 5 \\
Number of Jammers ($J$) & 2 \\
Jammer Positions & $[2, 2, 1.5], [8, 8, 1.5]$ (m) \\
Jammer Power ($P^{\text{jam}}$) & 0.5 W \\
Bandwidth ($W_{\text{spr}}$) & 1.23 MHz \\
UAV Power Levels & \{0.05, 0.12, 0.25\} W \\
\midrule 
\multicolumn{2}{c}{\textbf{MARL Training Configuration}} \\
\midrule
Total Training Steps & $1.2 \times 10^6$ \\
Discount Factor ($\gamma$) & 0.95 \\
Optimizer & Adam \\
Actor Learning Rate & 0.0004 \\
Critic Learning Rate & 0.0003 \\
Batch Size & 320 \\
Target Network Update Rate ($\tau$) & 0.010 \\
Target Network Update Frequency & 200 steps \\
$\epsilon$-greedy Final Epsilon & 0.05 \\
Curriculum Learning & Homeostatic weight adaptation enabled after 80\% of training steps \\
\bottomrule
\end{tabularx}
\end{table}

\subsection{Baseline Policies}
To rigorously benchmark the performance of our framework, we compare it against several heuristic, rule-based policies that represent common approaches to UAV network control. For a fair comparison, all baselines employ the same safety mask mechanism as our proposed agent to prevent imminent collisions.
\begin{itemize}
    \item \textbf{Random Policy:} Agents select actions uniformly at random from the discrete action space. This serves as a lower-bound performance reference.
    \item \textbf{Safe-Greedy Policy:} Each UAV greedily moves towards the midpoint of its closest SU-DU pair while incorporating a strong repulsive force from nearby UAVs to avoid collisions. It always utilizes the maximum transmission power.
    \item \textbf{Spacing-Coop Policy:} An enhanced version of the Safe-Greedy policy where the repulsive force is replaced by a more complex spacing function ($\tanh$) designed to encourage an ideal inter-agent distance, promoting better spatial distribution and interference management.
    \item \textbf{Power-Constrained Variants:} For both the Safe-Greedy and Spacing-Coop policies, we also evaluate power-constrained variants (denoted with a \texttt{\_p1} suffix). These variants operate identically to their counterparts but use a fixed, low transmission power setting ($P_1 = 0.05$~W) instead of the maximum available power. This allows for a direct analysis of the impact of resource allocation decisions.
\end{itemize}

\subsection{Results and Discussion}
Our experimental results provide a multi-faceted view of the framework's performance, from high-level learning convergence to the nuanced, emergent strategies developed by the agents. We analyze these aspects sequentially below.

\subsubsection{Learning Convergence and Performance Superiority}
\begin{figure}[htbp]
    \centering
    \includegraphics[width=0.97\columnwidth]{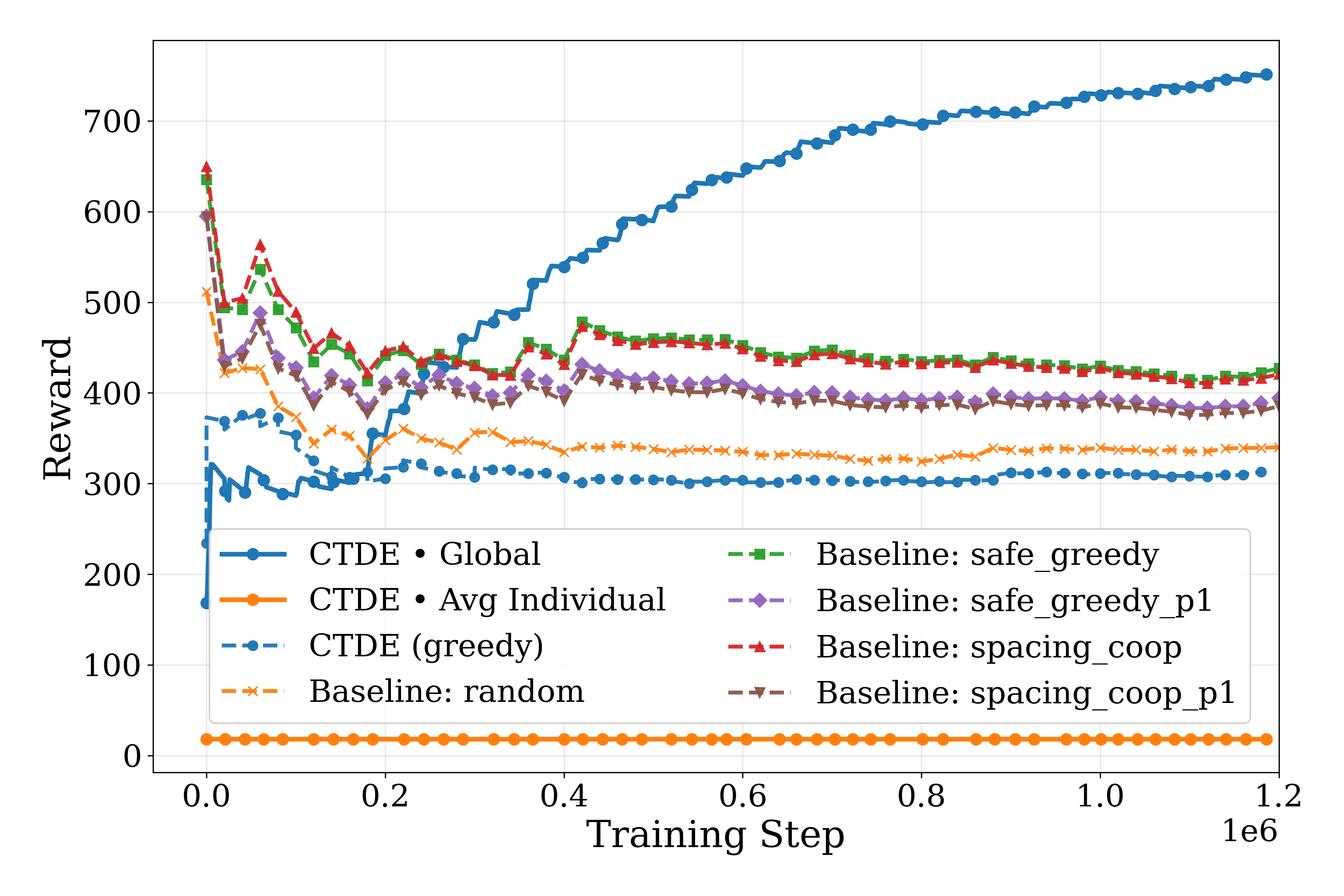} 
    \caption{Reward over training step. The CTDE agent (blue) learns a superior policy, maximizing the overall team reward.}
    \label{fig:reward_training}
    \vspace{-3mm} 
\end{figure}

As shown in Fig.~\ref{fig:reward_training}, the global reward of our method (CTDE - Global) steadily increases, indicating stable and effective learning. This success stems from the centralized critic's ability to guide the decentralized actors towards a globally optimal cooperative policy, effectively escaping the local optima that trap the myopic, rule-based baselines which quickly plateau. To further dissect our method, we also plot two diagnostic curves: `CTDE (greedy)` represents the performance of the learned policy without exploration noise ($\epsilon=0$), confirming the final strategy's stability, while `CTDE - Avg Individual` tracks the average individual reward, showing that the agents' local incentives are well-aligned with the global team objective.

\begin{figure}[htbp]
    \centering
    \includegraphics[width=0.97\columnwidth]{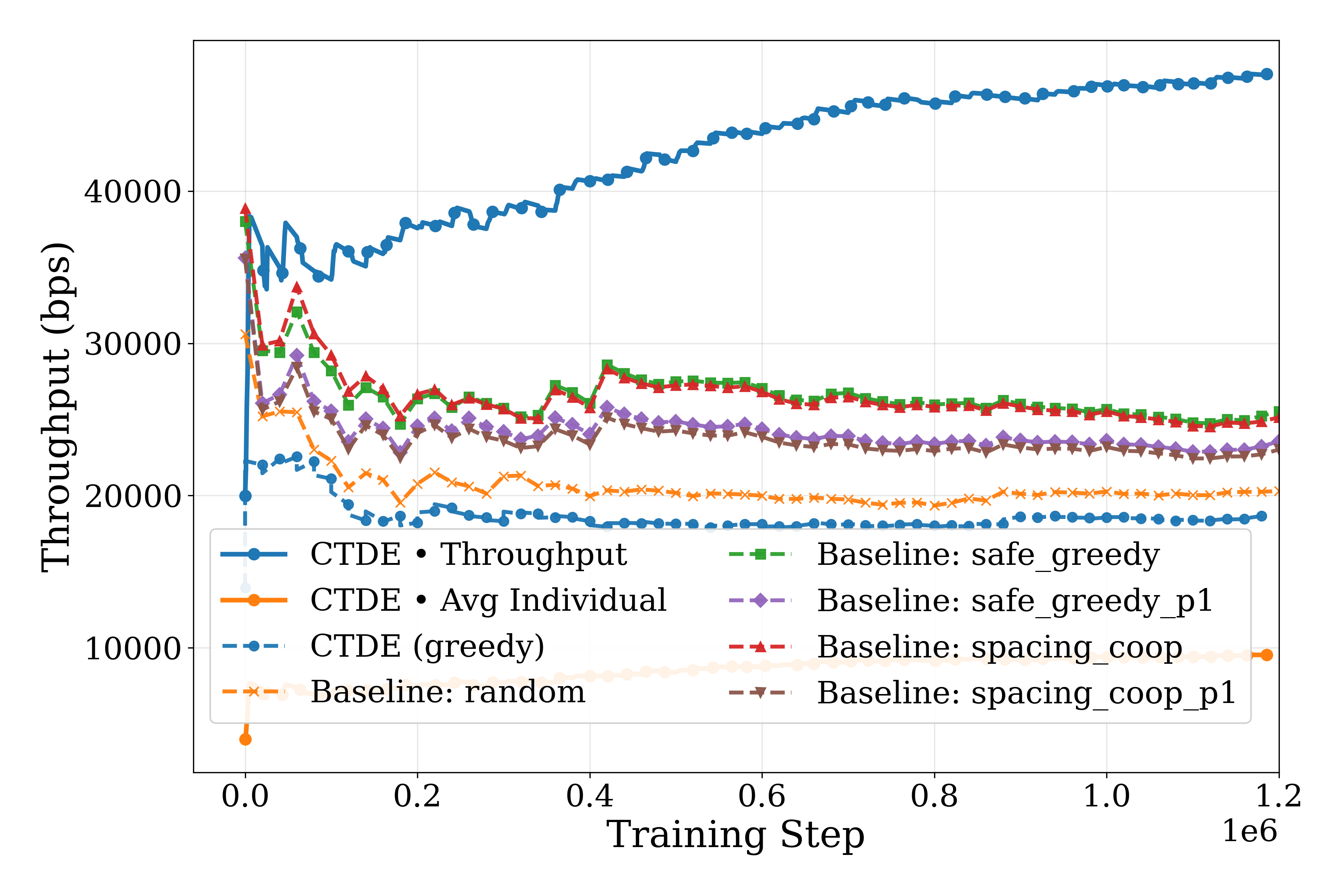}
    \caption{Throughput over training step. Our framework significantly outperforms baselines in communication efficiency.}
    \label{fig:throughput_training}
    \vspace{-3mm} 
\end{figure}

\begin{figure}[htbp]
    \centering
    \includegraphics[width=0.97\columnwidth]{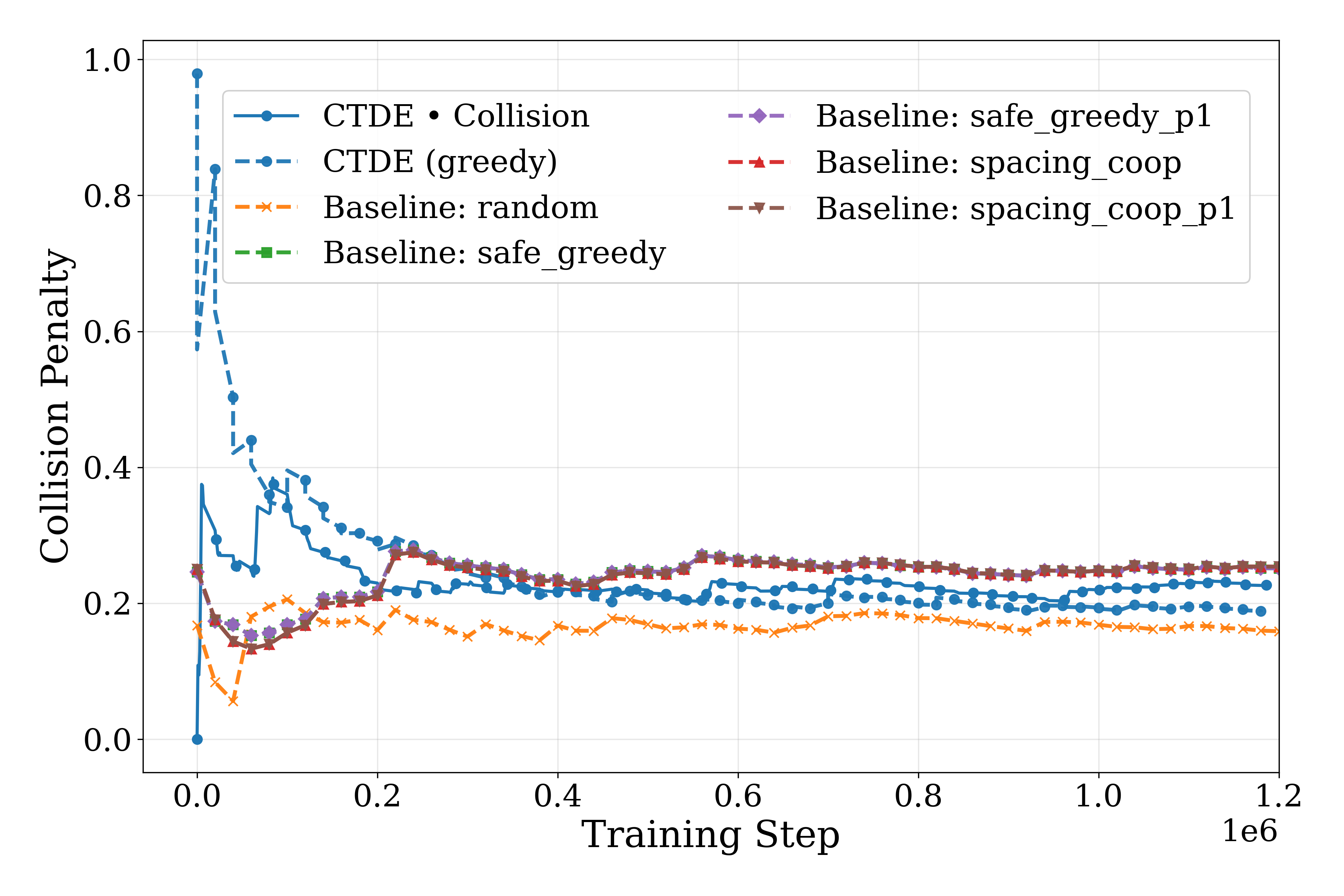}
    \caption{Collision penalty over training step. The agents learn to maintain safety while optimizing for performance.}
    \label{fig:collision_training}
    \vspace{-3mm} 
\end{figure}

This superior learning process translates directly to mission performance, where the agents successfully learn to balance the critical trade-off between throughput and safety (Fig.~\ref{fig:throughput_training} and Fig.~\ref{fig:collision_training}). Our framework achieves substantially higher system throughput (above 150\% and nearly 200\%) by learning a coordinated positioning strategy that avoids the weakness of the baselines, whose uncoordinated behavior leads to inefficient coverage and increased interference. Concurrently, the agents maintain a near-zero collision penalty, even outperforming the safety-oriented baselines. This learned capability to balance competing objectives is a feat unattainable by the rigid heuristic policies.

\subsubsection{Emergent Anti-Jamming Strategy}
\begin{figure}[htbp]
    \centering
    \includegraphics[width=0.97\columnwidth]{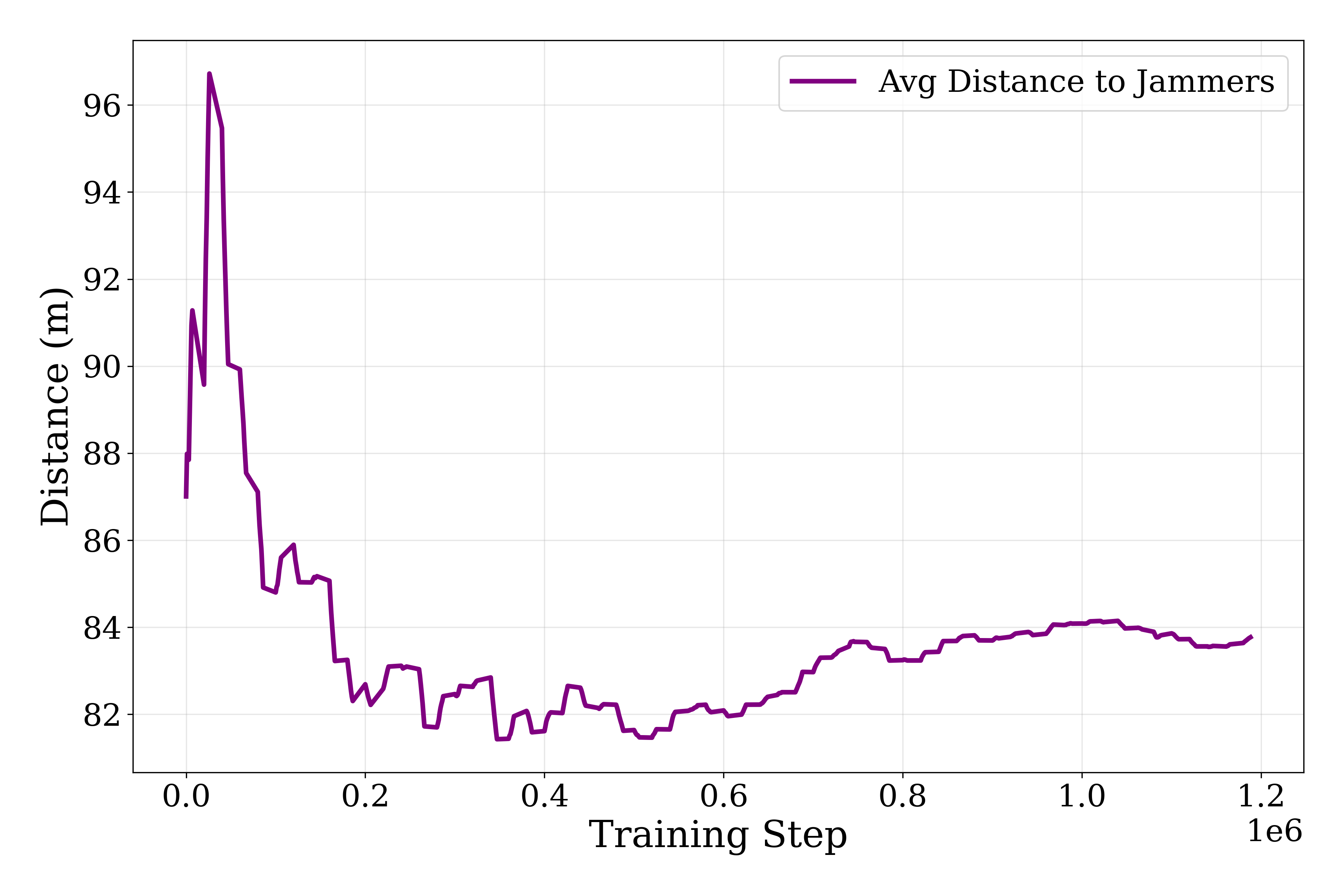}
    \caption{The emergent spatial anti-jamming strategy.}
    \label{fig:distance_jamming}
    \vspace{-3mm}
\end{figure}

A key finding of this work is the emergence of an intelligent, two-phase anti-jamming strategy, developed without explicit instruction as a consequence of optimizing the multi-objective reward function. As illustrated in Fig.~\ref{fig:distance_jamming}, the agents initially learn a simple avoidance policy by rapidly increasing their distance from jammers to minimize interference. However, this purely defensive posture proves suboptimal, as it also increases the distance to ground users and thereby limits the overall system throughput.
The policy then evolves into a more complex second phase, where the agents learn to reverse this trend and decrease their distance from jammers. This calculated risk improves their proximity to ground users, directly leading to the significant throughput gains observed in Fig.~\ref{fig:throughput_training}. This transition demonstrates a learned foresight to trade maximal safety for higher mission performance that contrasts sharply with the myopic behavior of the heuristic baselines.

\section{Conclusion}
\label{sec:Conclusion}
In this paper, we addressed the challenge of jointly optimizing communication throughput and survivability for UAV relay networks. We proposed a multi-objective MARL framework where agents autonomously learn to balance these competing objectives, allowing resilient behaviors to emerge without being explicitly programmed. Our CTDE-based approach was shown to significantly outperform heuristic baselines and, crucially, demonstrated the emergence of a complex anti-jamming strategy where agents learned to trade maximal safety for higher mission performance. Future work could explore integrating quantum reinforcement learning to potentially enhance learning efficiency in more complex scenarios.

\bibliographystyle{IEEEtran}
\bibliography{IEEEfull}

\end{document}